\documentclass[aps,prd,nofootinbib,twocolumn,preprintnumbers,superscriptaddress]{revtex4}
\pdfoutput=1
\RequirePackage{snapshot} 
\usepackage[utf8]{inputenc}
\usepackage{amsmath}
\usepackage{amssymb}
\usepackage{graphicx}
\usepackage{hyperref}
\usepackage{xspace}
\usepackage{wasysym} 
\usepackage{epstopdf}
\usepackage{booktabs}
\usepackage{hyperref}
\AtBeginDocument{
\heavyrulewidth=.08em
\lightrulewidth=.05em
\cmidrulewidth=.03em
\belowrulesep=.65ex
\belowbottomsep=0pt
\aboverulesep=.4ex
\abovetopsep=0pt
\cmidrulesep=\doublerulesep
\cmidrulekern=.5em
\defaultaddspace=.5em
}
\newcommand{\GeV}{\;\mathrm{GeV}}

\newcommand{\NNNLO}{\text{N$^3$LO}}

\usepackage{xcolor}
\definecolor{light-gray}{gray}{0.8}

\begin{document}

\title{Fully differential Vector-Boson Fusion Higgs Pair Production at Next-to-Next-to-Leading Order}

\preprint{OUTP-18-08P, ZU-TH 37/18}

\newcommand{\OXaff}{Rudolf Peierls Centre for Theoretical Physics,
  University of Oxford,\\
  Clarendon Laboratory, Parks Road, Oxford OX1 3PU}
\newcommand{\UZHaff}{Department of Physics, University of Z{\"u}rich, CH-8057 Z{\"u}rich, Switzerland}

\author{Fr\'ed\'eric A. Dreyer}
\affiliation{\OXaff}
\author{Alexander Karlberg}
\affiliation{\UZHaff}

\begin{abstract}
  We calculate the fully differential next-to-next-to-leading order (NNLO)
  QCD corrections to vector-boson fusion (VBF) Higgs pair production.
  This calculation is achieved in the limit in which there is no
  colored cross-talk between the colliding protons, using the
  projection-to-Born method.
  We present differential cross sections of key observables, showing
  corrections of up to $3-4\%$ at this order after typical VBF cuts,
  with the total cross section receiving contributions of about $2\%$.
  In contrast to single Higgs VBF production, we find that the NNLO
  corrections are for the most part within the next-to-leading order
  scale uncertainty bands.
\end{abstract}

\pacs{13.87.Ce,  13.87.Fh, 13.65.+i}

\maketitle

\section{Introduction}
Since the 2012 discovery of the Higgs
boson~\cite{Aad:2012tfa,Chatrchyan:2012xdj}, one of the focuses of the
experimental program of the Large Hadron Collider (LHC) is the
measurement of its couplings to fermions, to other bosons, and to
itself.
The self-coupling of the Higgs boson is of particular interest to
understand the electroweak symmetry breaking mechanism, and to
constrain new physics beyond the Standard Model (SM).
In that context, Higgs pair production will play a key role at the LHC~\cite{Aaboud:2018sfw,Aaboud:2018ewm,Aaboud:2018ftw,Aaboud:2018knk,Aaboud:2016xco,Aad:2015xja,Aad:2015uka,Aad:2014yja,Sirunyan:2018tki,Sirunyan:2018iwt,Sirunyan:2017guj,Sirunyan:2017djm,Sirunyan:2017tqo}, its
high luminosity upgrade (HL-LHC), and at future hadron colliders in
improving our understanding of the Higgs sector.

\begin{figure}[ht]
  \centering
  \includegraphics[width=0.6\linewidth]{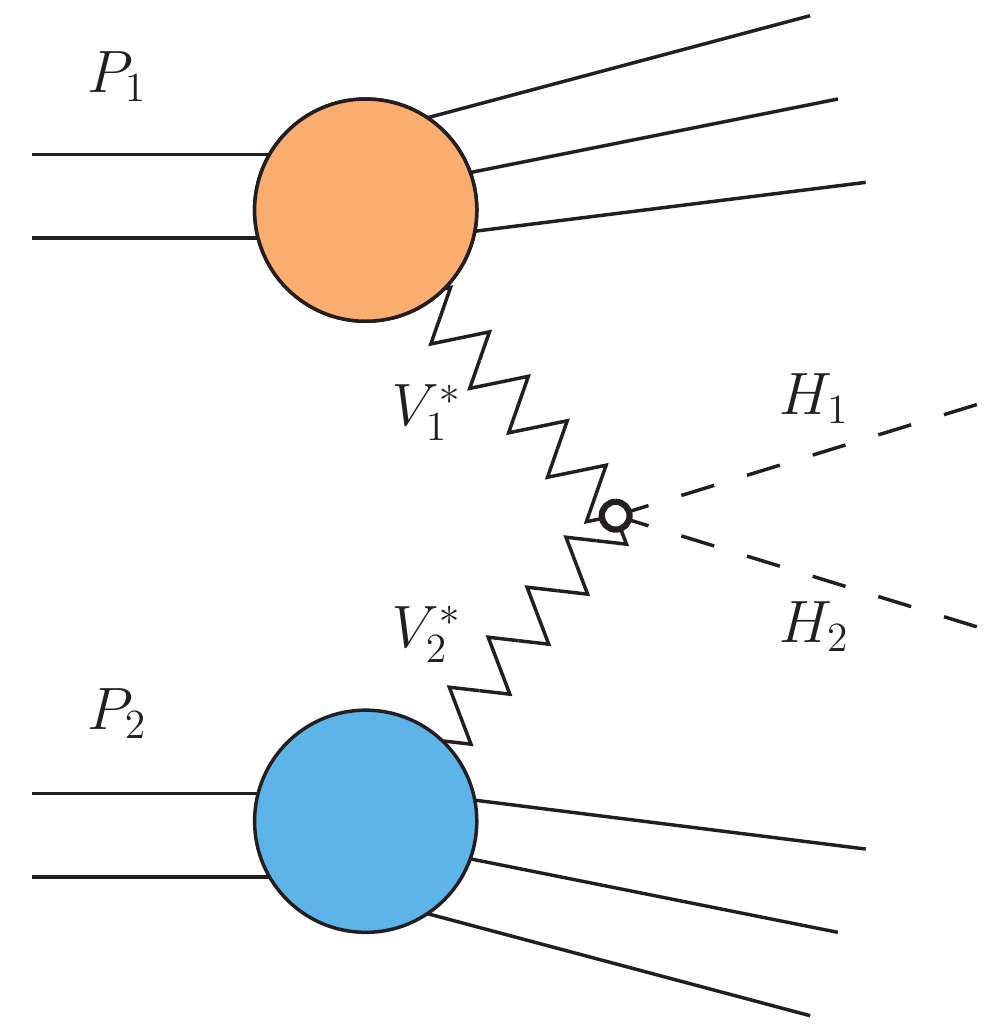}%
  \caption{ Diagram of VBF Higgs pair production, with the colored
    objects representing the incoming protons.  }
  \label{fig:vbfhh}
\end{figure}

In this article, we focus on double Higgs production via vector boson
fusion (VBF), shown in figure~\ref{fig:vbfhh}, which is the second
largest channel at the LHC after gluon-gluon fusion~\cite{deFlorian:2016spz}.
The VBF production mode is of particular interest in di-Higgs
production: due to the presence of two tagging jets, one can
significantly reduce the large backgrounds through an appropriate
choice of cuts; it also provides a unique sensitivity to deviations
from the SM in the trilinear Higgs coupling~\cite{Baglio:2012np}; and
is the most promising channel for measurements of the $hhVV$ quartic
coupling at the LHC~\cite{Bishara:2016kjn}.

Because of the important role that double Higgs production via VBF
will play at the LHC and beyond, it is crucial that precise
predictions for its production rate can be achieved.
So far the differential cross section has been calculated up to
next-to-leading order
(NLO)~\cite{Figy:2008zd,Baglio:2012np,Frederix:2014hta}, and up to
next-to-next-to-next-to-leading order (\NNNLO{}) in the structure function
approach~\cite{Han:1992hr} where all hadronic radiation is integrated
over~\cite{Liu-Sheng:2014gxa,Dreyer:2018qbw}.

Since cuts on the tagging jets are critical in reducing the large
experimental backgrounds, having access to the jet kinematics is
necessary to produce realistic predictions.
In particular, higher order calculations of single-Higgs VBF
production found that while corrections to the inclusive cross section
were small~\cite{Bolzoni:2010xr,Dreyer:2016oyx}, VBF cuts can
substantially impact the structure of these
contributions~\cite{Cacciari:2015jma,Cruz-Martinez:2018rod}.

We present for the first time the calculation of di-Higgs production
to NNLO in perturbative QCD for differential cross sections, which is
also the first of this type for a $2\to 4$ process.
This brings the VBF Higgs pair production channel to the same accuracy
as double Higgs gluon-gluon fusion
production~\cite{deFlorian:2013jea,deFlorian:2016uhr}.

These results are achieved using the deep inelastic scattering (DIS)
approximation, which corresponds to the limit in which there is no
colored cross-talk between the two colliding protons.
This approximation is exact at NLO, where the gluon exchange between the
protons is null for color reasons, and has been shown to be accurate
to more than 1\% at
NNLO~\cite{Ciccolini:2007ec,Harlander:2008xn,Bolzoni:2011cu} for
single Higgs VBF production.
Because the presence of an additional Higgs boson does not impact the
color flow between the hadrons, this limit is expected to be just as
accurate for Higgs pair production.
The approximation also corresponds to the exact calculation for two
sectors with different copies of QCD which interact purely through the
weak force, shown in different colors in figure~\ref{fig:vbfhh}.

\section{Calculation setup}

The VBF Higgs pair production cross section is calculated as a double
DIS process, and can be expressed
as~\cite{Han:1992hr}
\begin{align}
  \label{eq:vbfh-dsigma}
  d\sigma = &\,\sum_V\frac{2\,G_F^4 m_V^8}{s}
  \Delta_V^2(Q_1^2)
  \Delta_V^2(Q_2^2) \,
  d\Omega_{\text{VBF}} \notag
  \\
  &\times
    \mathcal{W}^V_{\mu\nu}(x_1,Q_1^2)
    \mathcal{M}^{V,\mu\rho}
    \mathcal{M}^{V*,\nu\sigma}
  \mathcal{W}^{V}_{\rho\sigma}(x_2,Q_2^2)\,.
\end{align}
Here $V= W^\pm, Z$ is the mediating boson, $G_F$ is Fermi's
constant, $m_V$ is the mass of the vector boson, $\sqrt{s}$ is the
collider center-of-mass energy, $\Delta_V^2$ is the squared boson
propagator, $Q_i^2=-q_i^2$ and $x_i = Q_i^2/(2 P_i\cdot q_i)$ are the
usual DIS variables, $\mathcal{W}^V_{\mu\nu}$ is the hadronic tensor
and $d\Omega_\text{VBF}$ is the four particle VBF phase space.
The matrix element of the vector boson fusion sub-process is given
by~\cite{Dobrovolskaya:1990kx}
%
\begin{widetext}
\begin{multline}
  \label{eq:VV-subproc}
  \mathcal{M}^{V,\mu\nu} = 
  \bigg[\bigg(1
  + \,\frac{4m_V^2}{(q_1 + k_1)^2 - m_V^2 + i\Gamma_V m_V}\,
  + \,\frac{6 \nu \lambda}{(k_1 + k_2)^2 - m_H^2 + i \Gamma_H m_H}
  \bigg)g^{\mu\nu}\, + \\
  + \,\frac{m_V^2}{(q_1 + k_1)^2 - m_V^2 + i \Gamma_V m_V}
  \frac{(2 k_1^\mu + q_1^\mu)(k_2^\nu - k_1^\nu - q_1^\nu)}{m_V^2 - i \Gamma_V m_V}
  \bigg] + (k_1 \leftrightarrow k_2)\,,
\end{multline}
\end{widetext}
%
where $k_1$ and $k_2$ are the Higgs momenta such that
$q_1 + q_2 = k_1 + k_2$, $\lambda$ is the SM trilinear Higgs
self-coupling, $\nu$ is the vacuum expectation value of the Higgs
field, and we have absorbed a factor $\sqrt{2}\, G_F m_V^2$ from
$\mathcal{M}^V$ into the overall normalization of the cross section.

To obtain differential results, we use the projection-to-Born
method~\cite{Cacciari:2015jma}, combining an inclusive NNLO
calculation with a calculation at NLO of two Higgs bosons in association
with three jets.
The electroweak di-Higgs three jet process was obtained by modifying
the matrix elements in \texttt{POWHEG}'s corresponding single-Higgs
process~\cite{Alioli:2010xd,Jager:2014vna}.
We cross-checked this implementation by verifying that the kinematical
properties of the third and fourth jets agreed within statistical
uncertainties with predictions obtained through the
\texttt{MadGraph5\_aMC@NLO}~\cite{Alwall:2014hca} framework.
We also compared our results at NLO with
\texttt{VBFNLO}~\cite{Arnold:2011wj}, finding agreement at the few
permille level.

For our predictions, we consider 14 TeV proton-proton collisions, with
the \texttt{PDF4LHC15\_nnlo\_mc}~\cite{Butterworth:2015oua} set as
implemented in \texttt{LHAPDF} version 6.1.6~\cite{Buckley:2014ana}, a
Higgs mass $m_H = 125\GeV$ and electroweak parameters fixed by
$m_W=80.379\GeV$, $m_Z=91.1876\GeV$, and
$G_F=1.16637 \cdot 10^{-5} \GeV^{-2}$ consistent with the most recent
Review of Particle Physics~\cite{Tanabashi:2018oca}. For internal
propagators we use $\Gamma_Z=2.4952\GeV$, $\Gamma_W=2.141\GeV$, and
$\Gamma_H=4.030\cdot10^{-3}\GeV$, while the final state Higgs bosons are
considered in the narrow-width approximation.
The central scale is chosen analogously to the one considered in the
single-Higgs NNLO calculation~\cite{Cacciari:2015jma}
\begin{equation}
  \label{eq:scale}
  \mu_0^2(p_{t,HH}) = \frac{m_H}{2} \sqrt{\left(\frac{m_H}{2}\right)^2 +
        p_{t,HH}^2}\,,
\end{equation}
motivated by the fact that it approximates $\sqrt{Q_1^2Q_2^2}$
relatively well.
The uncertainties from missing higher orders are estimated by
simultaneously varying the renormalization and factorization scale up
and down by a factor two.
%
\begin{table}[t] 
  \centering
  \phantom{x}\medskip
  \begin{tabular}{lcccc}
    \toprule
    &&  $\sigma^\text{(no cuts)}$  [fb]  && $\sigma^\text{(VBF cuts)}$ [fb] \\
    \midrule
    LO       &&  $2.016\,^{+0.164}_{-0.142}$    &&  $0.799\,^{+0.082}_{-0.069}$\\[4pt]
    NLO      &&  $2.049\,^{+0.007}_{-0.021}$    &&  $0.726\,^{+0.005}_{-0.020}$\\[4pt]
    NNLO     &&  $2.053\,^{+0.000}_{-0.003}$    &&  $0.713\,^{+0.004}_{-0.001}$\\
    \bottomrule
  \end{tabular}
  \caption{Cross sections at LO, NLO and NNLO for VBF Higgs
    production, fully inclusively and with VBF cuts.  The
    uncertainties are obtained from a three-point scale variation. The
    statistical uncertainty on the inclusive and fiducial cross
    sections are below one permille.}
\label{tab:cross-sections}
\end{table}

The events have to pass a set of VBF selection cuts. We require at least two hard jets with
\begin{align}
  p_{t,j}> 25 \GeV, \quad |y_j|< 4.5.
\end{align}
The two hardest jets additionally have to satisfy
\begin{align}
  m_{j_1,j_2} > 600 \GeV, \quad |y_{j_1}-y_{j_2}| > 4.5, \quad y_{j_1}y_{j2}<0.
\end{align}
We define our jets using the anti-$k_t$
algorithm~\cite{Cacciari:2008gp} as implemented in \texttt{FastJet} version
3.3.0~\cite{Cacciari:2011ma}.
The jet radius is set to $R=0.4$.

\begin{figure*}
  \centering
          \includegraphics[clip,width=0.25\textwidth,page=3,angle=0]{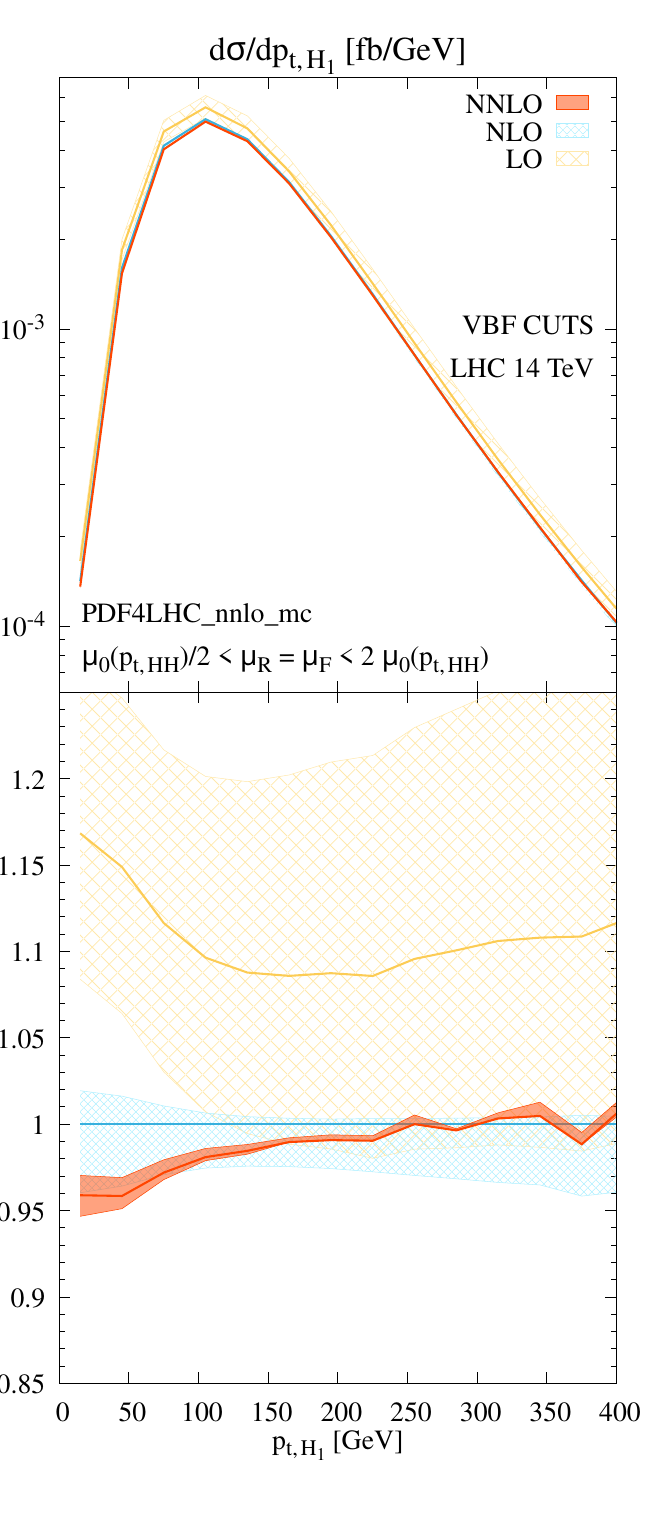}%
    \hfill\includegraphics[clip,width=0.25\textwidth,page=4,angle=0]{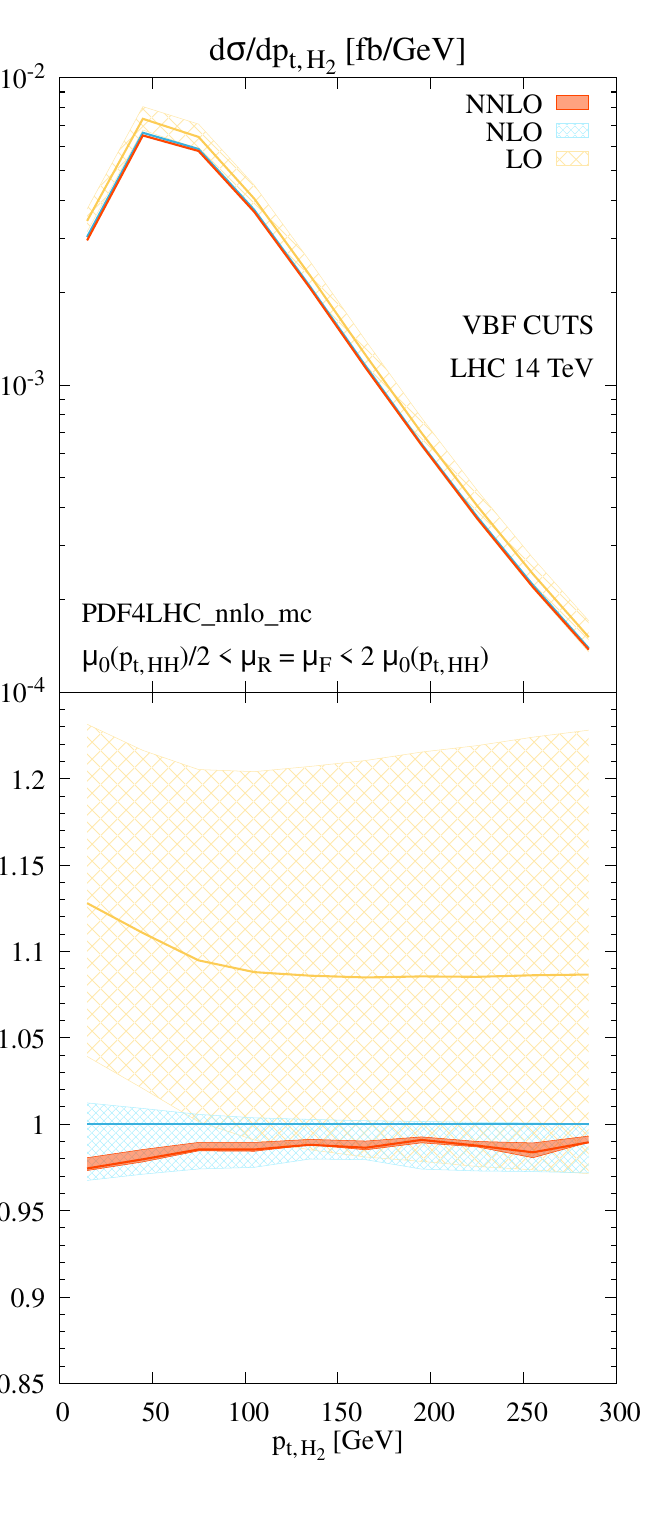}%
    \hfill\includegraphics[clip,width=0.25\textwidth,page=5,angle=0]{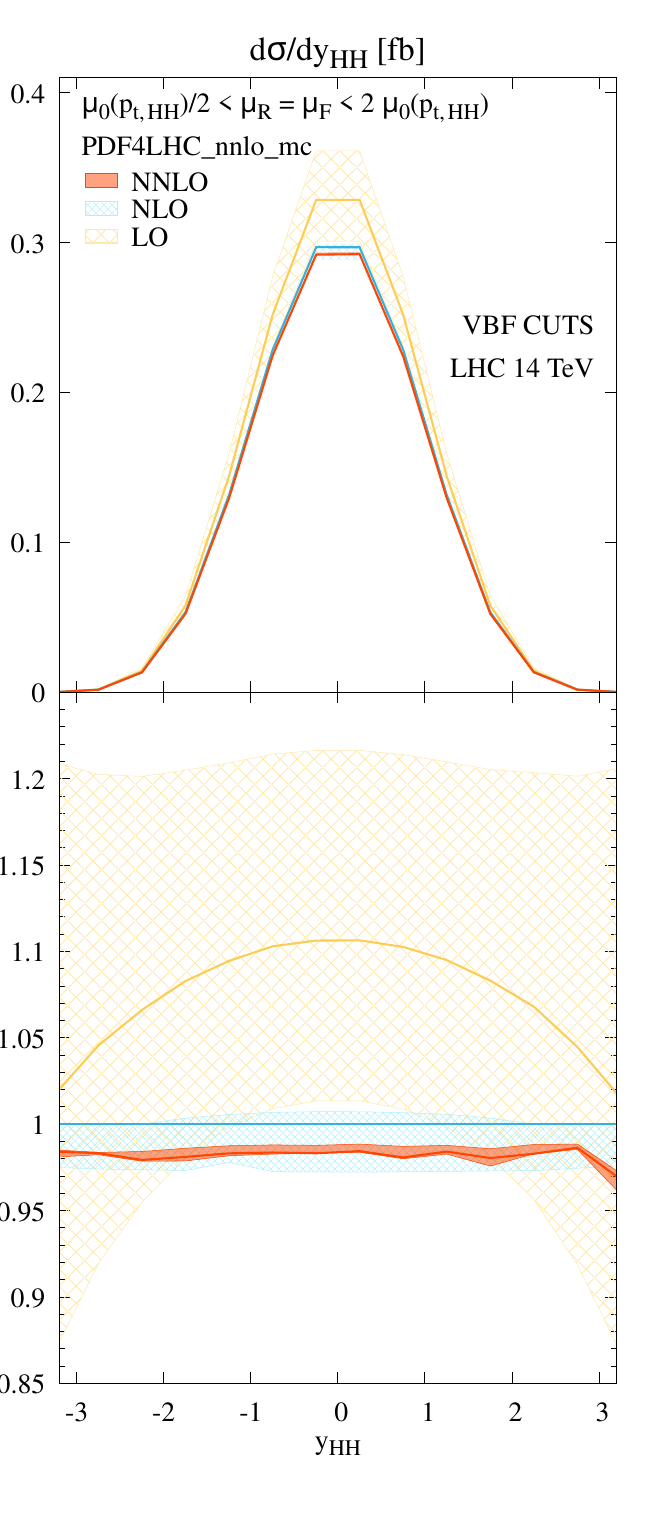}%
    \hfill\includegraphics[clip,width=0.25\textwidth,page=6,angle=0]{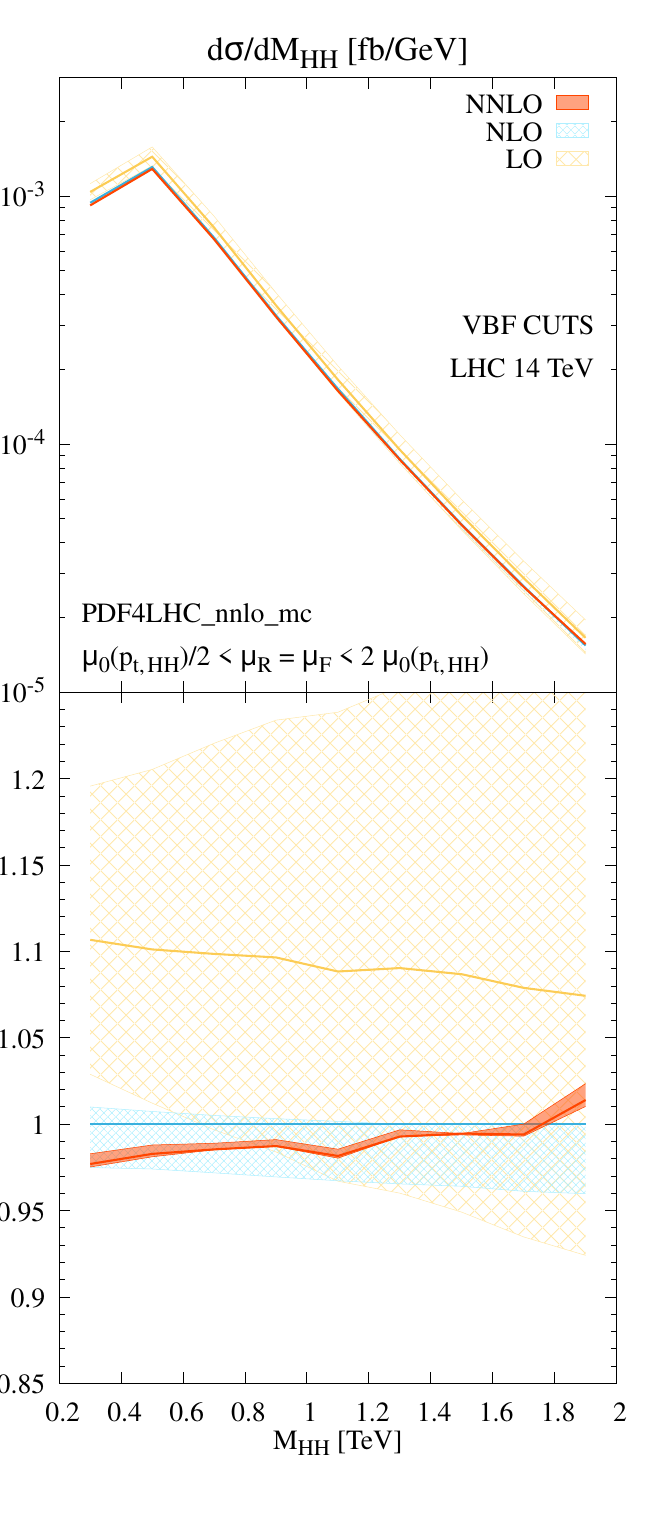}%
    \caption{From left to right, differential cross sections for the
      transverse momentum distributions for the two Higgs bosons,
      $p_{t,H_1}$ and $p_{t,H_2}$, and the distribution of the
      rapidity and invariant mass of the Higgs pair, $y_{HH}$ and
      $m_{HH}$.
      The lower panel shows the ratio to the NLO prediction.
    }
  \label{fig:diff-cross-sections-higgs}
\end{figure*}

\section{Results}

The total cross sections before and after VBF cuts are given in
table~\ref{tab:cross-sections}.
The errors given correspond to the envelope of the three-point scale
variation, while the statistical uncertainties are negligible.
We can observe here that the convergence of the inclusive cross
section is better than that of the fiducial cross section, with a
second order correction of only $2\permil$, while it is of about
$1.7\%$ after cuts.
At NNLO, the scale variation uncertainty on the fiducial cross section
is reduced by a factor four, to less than $1\%$.
This stands in contrast to the single-Higgs process, where there
is almost no reduction in scale uncertainties at NNLO, and where the NLO
variation bands do not encompass the central value of the NNLO
prediction.
One reason for this can likely be found in the different kinematics
probed in di-Higgs VBF production. At leading order the two tagging
jets take the full recoil of the two fusing vector bosons. Since twice
as much energy is required to produce two Higgs bosons compared to
one, we expect the jets in di-Higgs VBF production to be harder than
the jets in single Higgs VBF production. For that reason more events
will pass the VBF cuts even when they radiate outside of the jet cone.

In figure~\ref{fig:diff-cross-sections-higgs}, we present the
differential distributions after VBF cuts of the Higgs bosons.
The Higgs bosons are ordered in transverse momentum, with the harder
one labeled $H_1$, and the softer boson $H_2$.
We show the transverse momentum of both Higgs bosons, $p_{t,H_1}$ and
$p_{t,H_2}$, as well as the rapidity and invariant mass of the Higgs
pair, $y_{HH}$ and $m_{HH}$.
We see that the NNLO corrections can be sizeable, at about $4\%$, in
the region where both Higgs bosons are soft, while for Higgs bosons with large
transverse momentum the corrections become quite small, in the $1-2\%$
range.
The invariant mass of the Higgs pair is an observable of particular
interest, as its distribution can be very sensitive to deviations from
the SM in certain models~\cite{Kaplan:1983fs,Contino:2010mh,Bishara:2016kjn}.
Here we can observe moderate NNLO corrections, up to about $2\%$,
albeit with some kinematical dependence.

\begin{figure*}
  \centering
          \includegraphics[clip,width=0.25\textwidth,page=1,angle=0]{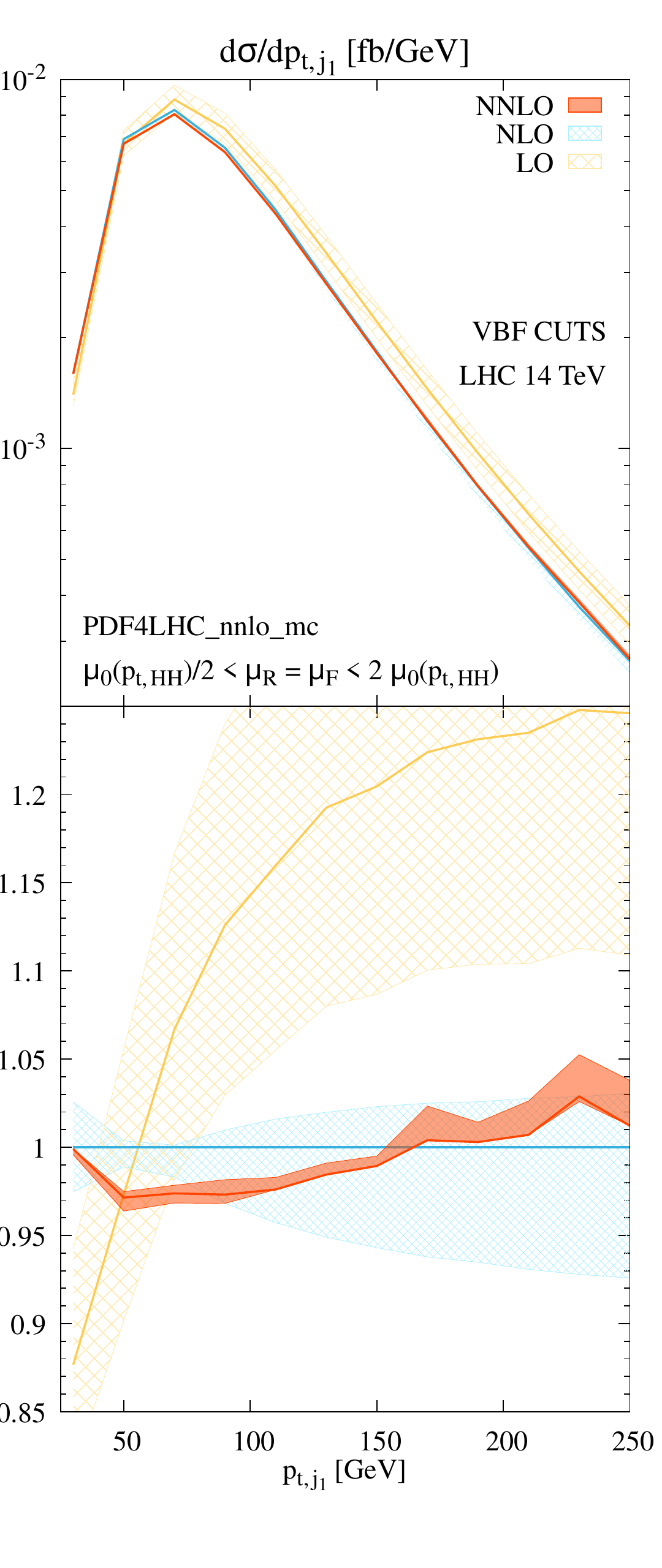}%
    \hfill\includegraphics[clip,width=0.25\textwidth,page=2,angle=0]{vbfhh.pdf}%
    \hfill\includegraphics[clip,width=0.25\textwidth,page=9,angle=0]{vbfhh.pdf}%
    \hfill\includegraphics[clip,width=0.25\textwidth,page=8,angle=0]{vbfhh.pdf}%
    \caption{From left to right, differential cross sections for the
      transverse momentum distributions for the two leading jets,
      $p_{t,j_1}$ and $p_{t,j_2}$, and the distribution for the
      rapidity and azimuth separation between the two leading jets,
      $\Delta y_{j_1,j_2}$ and $m_{j_1,j_2}$.
      The lower panel shows the ratio to the NLO prediction.}
  \label{fig:diff-cross-sections-jets}
\end{figure*}

Differential cross sections after VBF cuts for the tagging jets are shown in
Fig.~\ref{fig:diff-cross-sections-jets}.
Here it is interesting to see that the structure of the NNLO
corrections is quite different than in the single-Higgs process.
In particular, they are as for the fiducial cross section smaller and
mostly contained within the scale variation bands of the NLO
predictions.
The transverse momentum of the two tagging jets receives corrections
of $2-4\%$, with the corrections being more pronounced at moderate
$p_t$ values.
For the rapidity separation between the jets and their invariant mass,
the corrections are of about $2\%$, with little dependence on the
observable, except at small values of the rapidity separation, where
the corrections become negligible.
%


As we pointed out earlier, the tagging jets are expected to be harder
than in single-Higgs production, a fact which is confirmed by
comparing the slopes of the two leading jets for both processes.
We find that in single-Higgs production the slope of the transverse
momentum of the two hardest jets falls more steeply.
This partly explains the difference in the structure of the higher
order QCD corrections between single- and di-Higgs production.
%

\section{Other contributions}

We note that there are theoretical effects beyond the perturbative QCD
corrections considered in this article.
We have estimated the contribution of the $s$-channel, which is not
included in the DIS approximation using \texttt{MadGraph5\_aMC@NLO}
version 2.6.2.
While it has a substantial effect on the inclusive cross section, of
about $27\%$, it is reduced to one permille after our VBF cuts are
applied and can thus be neglected.
There are also non-factorizable contributions to VBF which first
appear at NNLO and are neglected in the DIS approximation. These
should, as in the case of single Higgs production, contribute to less
than one percent of the cross
section~\cite{Ciccolini:2007ec,Harlander:2008xn,Bolzoni:2011cu}.

There are also corrections due to higher order electroweak
effects.
They are currently unknown, but can be estimated from dominant light
quark induced channels using
\texttt{Recola(Collier)+MoCaNLO}~\cite{Actis:2016mpe,Denner:2016kdg,MoCaNLO,Bendavid:2018nar}
for the di-Higgs and single-Higgs VBF process, comparing the latter to
\texttt{HAWK} ~\cite{Denner:2014cla}.
For di-Higgs VBF production the EW corrections lie between $-5\%$ and
$-7\%$ for both the inclusive and fiducial cross sections.
Compared to the fiducial single-Higgs VBF prediction of roughly
$-8.5\%$ (using the same set-up, i.e. excluding photonic and b-quark
channels) the di-Higgs VBF process thus does not seem to receive large
VBS-like corrections~\cite{Biedermann:2016yds,Biedermann:2017bss}.
The full electroweak corrections should therefore contribute at about
the same level as the NNLO QCD corrections reported here.

Cross sections after cuts are also affected by non-perturbative effects.
Using \texttt{Pythia} version 8.219~\cite{Sjostrand:2014zea} with the
default 4C tune, we evaluate the correction to the fiducial cross
section due to hadronization to about $-1\%$, while underlying event
leads to a change of $+4\%$ to the cross section after cuts.

Finally, there are uncertainties associated with the determination of
parton distribution functions and the strong coupling constant.
Using the \texttt{PDF4LHC\_nnlo\_mc\_pdfas} set, we find these to be
of $2.1\%$ on the inclusive cross section.

\section{Conclusions}

In this paper we presented the first differential calculation of NNLO
QCD corrections to VBF Higgs pair production.
The calculation is fully differential in the kinematics of the tagging
jets, allowing for precise predictions of fiducial cross sections.
This was achieved by using the projection-to-Born method, combining an
inclusive NNLO calculation with a differential NLO calculation with an
additional final state jet.
We showed that after typical VBF cuts, the NNLO corrections are
moderate, with corrections of $1.7\%$ to the total cross section and
up to $4\%$ in some differential observables.
The corrections can have a non-trivial kinematic dependence, and are
notably different from the single-Higgs production mode.

The full code used for this calculation is available as part of
\href{https://provbfh.hepforge.org/}{\texttt{proVBFHH}
  v1.0.0}~\cite{proVBFHH}, along with a lightweight program,
\href{https://provbfh.hepforge.org/}{\texttt{proVBFH-incl} v2.0.0},
which allows for the calculation of inclusive cross sections up to
\NNNLO{} for both single and double Higgs production.

\textbf{Acknowledgments:} 
We thank Michael Rauch for collaboration in the early stages of this
project and helpful feedback on \texttt{VBFNLO}, and Marc Zaro for
discussions about \texttt{MadGraph5\_aMC@NLO}. We are grateful to
Jean-Nicolas Lang and Mathieu Pellen for providing us with an estimate
of the electroweak corrections.
We also wish to thank Gavin Salam, Giulia Zanderighi and Markus Ebert
for valuable comments on the manuscript.
F.D.\ thanks the University of Zurich and the Pauli Center for
Theoretical Studies, and A.K.\ thanks the University of Oxford and the
Rudolf Peierls Centre for Theoretical Physics for hospitality while
this work was being completed.
F.D.\ is supported by the Science and Technology Facilities Council
(STFC) under grant ST/P000770/1.
A.K.\ is supported by the Swiss National Science Foundation (SNF)
under grant number 200020-175595.

\bibliography{dihiggs}

\end{document}